\documentclass[12pt,english]{article}
\usepackage[LGR,T1]{fontenc}
\usepackage[latin9]{inputenc}
\usepackage[a4paper]{geometry}
\usepackage{textcomp}
\usepackage{amsmath}
\usepackage{graphicx}
\usepackage{setspace}
\usepackage{wasysym}
\usepackage{esint}
\usepackage{subscript}
\makeatletter
\@ifundefined{date}{}{\date{}}
\usepackage{babel}
\usepackage{color}

\newcommand{\FeO}{$\alpha$-Fe$_2$O$_3$}
\newcommand{\RhFeO}{$\alpha$-Fe$_{1.97}$Rh$_{0.03}$O$_3$}
\usepackage[superscript,nomove,biblabel]{cite}
\usepackage{times}
\usepackage{lineno}

\makeatother

\usepackage{babel}
\begin{document}
\begin{center}
\textbf{\large{}Antiferromagnetic Half-skyrmions and Bimerons at room
temperature}{\large\par}
\par\end{center}

\begin{center}
\ \ 
\par\end{center}

\begin{center}
Hariom Jani$^{1*}$, Jheng-Cyuan Lin$^{2}$, Jiahao Chen$^{2}$, Jack
Harrison$^{2}$, Francesco Maccherozzi$^{3}$, Jonathon Schad$^{4}$,
Saurav Prakash$^{1}$, Chang-Beom Eom$^{4,5}$, A. Ariando$^{1,6}$,
T. Venkatesan$^{7*}$, Paolo G. Radaelli$^{2*}$ 
\par\end{center}

\begin{center}
\ \ 
\par\end{center}

\begin{center}
$^{1}$Department of Physics, National University of Singapore, Singapore 
\par\end{center}

\begin{center}
$^{2}$Clarendon Laboratory, Department of Physics, University of
Oxford, UK 
\par\end{center}

\begin{center}
$^{3}$Diamond Light Source, Harwell Science and Innovation Campus,
UK 
\par\end{center}

\begin{center}
$^{4}$Department of Materials Science and Engineering, University
of Wisconsin-Madison, USA 
\par\end{center}

\begin{center}
$^{5}$Department of Physics, University of Wisconsin-Madison, USA 
\par\end{center}

\begin{center}
$^{6}$NUS Graduate School for Integrative Sciences and Engineering,
National University of Singapore, Singapore 
\par\end{center}

\begin{center}
$^{7}$Department of Electrical and Computer Engineering, National
University of Singapore, Singapore 
\par\end{center}

\begin{center}
\ \ 
\par\end{center}

\begin{center}
$^{*}$Correspondence to: hariom.k.jani@u.nus.edu, venky@nus.edu.sg,
paolo.radaelli@physics.ox.ac.uk
\par\end{center}

\pagebreak{}

\textbf{In the quest for post-CMOS technologies, ferromagnetic skyrmions}
\textbf{and their anti-particles have shown great promise as topologically
protected \cite{2020_roadmap,Tokura_Frustrated_magnet_skyrmions,Geoff_RTskyrmions_and_current_driven_motion,FM_Skyrmion_2,FM_skyrmion_3,Anjan_skyrmions,Parkin_antiskrymions,meron_skyrmion_lattice}
solitonic information carriers in memory-in-logic or neuromorphic
devices \cite{2020_roadmap,Skyrmions_application,Skyrmion_electronics,Neuromorphic_spintronics}.
However, the presence of dipolar fields in ferromagnets, restricting
the formation of ultra-small topological textures \cite{Skyrmions_application,Geoff_RTskyrmions_and_current_driven_motion,Anjan_skyrmions,meron_skyrmion_lattice,SkHE_of_skyrmions},
and the deleterious skyrmion Hall effect when driven by spin torques
\cite{SkHE_of_skyrmions,Skyrmions_application,Skyrmion_electronics}
have thus far inhibited their practical implementations. Antiferromagnetic
analogues, which are predicted to demonstrate relativistic dynamics,
fast deflection-free motion and size scaling have recently come into
intense focus \cite{AFM_skyrmion_predict_1,SAF_skyrmions_no_SkHE_NatComm_Theory,Antiferromagnetic_spintronics_RMP,Skyrmions_application,Current_driven_motion_AFM_bimerons_PRL,Ferrimagnet_fastmotion,Ohno-SAF_skyrmions_no_SkHE,SAF_skyrmion},
but their experimental realizations in natural antiferromagnetic systems
are yet to emerge. Here, we demonstrate a family of topological antiferromagnetic
spin-textures in $\boldsymbol{\alpha}$-Fe$\boldsymbol{_{2}}$O$\boldsymbol{_{3}}$
-- an earth-abundant oxide insulator -- capped with a Pt over-layer.
By exploiting a first-order analogue of the Kibble-Zurek mechanism
\cite{Kibble1976,Zurek1985}, we stabilize exotic merons-antimerons
(half-skyrmions\cite{meron_skyrmion_lattice}), and bimerons\cite{Bimerons_1,Current_driven_motion_AFM_bimerons_PRL},
which can be erased by magnetic fields and re-generated by temperature
cycling. These structures have characteristic sizes of the order $\sim$100
nm that can be chemically controlled via precise tuning of the exchange
and anisotropy, with pathways to further scaling. Driven by current-based
spin torques from the heavy-metal over-layer, some of these antiferromagnetic
textures could emerge as prime candidates for low-energy antiferromagnetic
spintronics at room temperature \cite{2020_roadmap,Skyrmion_electronics,Skyrmions_application,Neuromorphic_spintronics,AFM_skyrmion_Logic_gates}. }

Topological textures in antiferromagnetic (AFM) materials, consisting
of compensated ferromagnetic (FM) sublattices, produce negligible
stray fields. Nonetheless, the order parameter of AFMs $-$ the Néel
vector ($\boldsymbol{L}$) $-$ twists in these textures so as to
generate a non-zero \emph{Néel topological charge} ($Q$) \cite{AFM_skyrmion_predict_1}.
This property endows AFM textures with topological protection, while
alleviating the limitations encountered by their FM counterparts \cite{AFM_skyrmion_predict_1,Antiferromagnetic_spintronics_RMP,Skyrmions_application,SAF_skyrmions_no_SkHE_NatComm_Theory,2020_roadmap}.
Furthermore, insulating AFMs are predicted to host ultra-fast spin
dynamics and relativistic motion of textures up to \textasciitilde{}
few km/s \cite{Antiferromagnetic_spintronics_RMP,NiO_low_damping,Galkina,AFM_skyrmion_predict_1,Current_driven_motion_AFM_bimerons_PRL}\ due
to low Gilbert damping \cite{Klaui_Long_distance_transport_Fe2O3,NiO_low_damping,AFM_skyrmion_predict_1}\ and
the exchange amplification effect \cite{Galkina}. Lastly, due to
negligible Joule losses during magnon-mediated spin transport, AFM
insulators are also important candidates for low-power spintronics
\cite{Klaui_Long_distance_transport_Fe2O3,AFMI_magnon_torque}. Therefore,
generating and controlling AFM topological textures in insulators
is crucial both for their fundamental understanding and for applications.

The latest experimental observations of skyrmions in metal-based synthetic
AFM stacks have exhibited some of the desirable properties\cite{SAF_skyrmion,Ohno-SAF_skyrmions_no_SkHE};
however, equivalent demonstrations in a natural AFM material have
been absent. Recently, we discovered that the insulating AFM oxide
\FeO\ (hematite), when interfaced with a ferromagnetic over-layer
(Co), can support flat (anti)vortices in the AFM layer \cite{Francis_Fe2O3_vortex}.
However, these AFM textures, which are \emph{not} equivalent to skyrmions,
were nucleated during an irreversible process at sample growth and
are unsuitable for some applications due to the inability to provide
spin-orbit torques (SOTs) directly at the AFM interface. In contrast,
here we demonstrate a variety of AFM topological textures, including
analogues of skyrmions, at the interface between \FeO\ and Pt. Furthermore,
by varying temperature and doping we show field-free and reversible
stabilization of these AFM textures, which can be nucleated below
or at room temperature by simple chemical tuning.

\FeO\ crystallizes in the trigonal corundum structure (space group:
$R\overline{3}c$), with alternating ferromagnetic sublattices ($\boldsymbol{M}_{\boldsymbol{1,2}}$)
stacked anti-parallel along the $c$-axis, Fig. 1a (insets). Its spin
orientation is governed by the axial magneto-crystalline anisotropy
$K(T)$, which varies with temperature due to the delicate interplay
between magnetic-dipolar and single-ion interactions. \FeO\ is one
of the few systems where $K(T)$ undergoes a sign reversal, causing
the spins to reorient across the Morin transition temperature $T_{M}$,
which can be tuned by chemical substitutions \cite{Besser_dopedFe2O3,Coey_Rh-doped_Fe2O3}.
The Néel vector, $\boldsymbol{L}\equiv\boldsymbol{M_{1}-M_{2}}$,
lies \emph{in plane} (IP) above $T_{M}$ and flips \emph{out of plane}
(OOP) below $T_{M}$. Above $T_{M}$, the presence of a much weaker
basal anisotropy term \cite{Besser_dopedFe2O3} results in the formation
of six distinct IP $\boldsymbol{L}$-domains, \emph{i.e.} three pairs
of directional domains, each consisting of elements related by time
reversal. For our thin-film \FeO\ samples, grown epitaxially on
$(0001)$-oriented $\alpha$-Al\textsubscript{2}O\textsubscript{3}
substrates by pulsed laser deposition (Methods, Extended Data Fig.
1), the Morin transition occurs at $T_{M}\sim$ 240 K. We also grew
Rh-substituted films with the composition \RhFeO, which display an
elevated Morin temperature $T_{M}\sim$ 298 K (Extended Data Fig.
3) \cite{Coey_Rh-doped_Fe2O3}. All our films for the photo-emission
electron microscopy (PEEM) experiments were capped with a 1-nm Pt
over-layer to prevent charging. Pt does not change the magnetic properties
of the films, and is ideally suited for spintronic applications, due
to its large spin-orbit coupling. The $\boldsymbol{L}$ reorientation
across the respective $T_{M}$ for these samples was confirmed by
X-ray magnetic linear dichroism (XMLD), resonant at the Fe-L edge
energies, Extended Data Fig. 1. Due to the \emph{bulk} Dzyaloshinskii-Moriya
interaction (DMI), when $T>T_{M}$ a small IP canting ($\sim$1.1
mrad) occurs between the sublattices $\boldsymbol{M}_{\boldsymbol{1,2}}$,
which can be observed as a small hysteretic IP moment ($\boldsymbol{m=\boldsymbol{M_{1}+M_{2}}}$),
Fig. 1a and Extended Data Fig. 2.

To map the spatial distribution of the AFM textures, we performed
XMLD-PEEM experiments as a function of temperature. Differential energy-contrast
images, produced by mapping the quantity $\Delta=(I_{E_{1}}-I_{E_{2}})/(I_{E_{1}}+I_{E_{2}})$,
were collected using both vertical (LV) and horizontal (LH) linear
polarization of the X-rays, incident at a 16° grazing angle onto the
films, with the LH polarization axis parallel to the sample plane.
The energies $E_{1}$ and $E_{2}$ are in the shoulders of the Fe-L\textsubscript{\emph{III}}
double peaks, and were chosen as they are most sensitive to AFM XMLD
contrast (Methods) \cite{Elke_Fe3O4_Fe_edge,Francis_Fe2O3_vortex}.
With this technique, we can clearly distinguish between the OOP and
any IP spin orientations, since the quantity $\Delta$ has opposite
dichroism in the two cases. Moreover, we can combine LH images collected
at several different sample azimuthal angles (we typically use 6 angles
\cite{Francis_Fe2O3_vortex}) to reconstruct an IP Néel vector map
(phase image of $\boldsymbol{L}$), with the caveat that the sign
of $\boldsymbol{L}$ cannot be distinguished by XMLD (see Methods
and Supplementary Section 1). In spite of this ambiguity, we can clearly
determine the topological character of most spin textures through
their Néel winding numbers $w$ and the \emph{magnitude} of their
Néel topological charges $|Q|$. The \emph{sign} of $Q$ cannot be
determined by XMLD but, in a layered AFM, this is only defined with
respect to an arbitrary reference layer (see Methods).

\subsection*{Evolution across the Morin transition}

Fig. 1b-1l show LV-PEEM images of the \FeO$-$Pt interface at different
temperatures while warming through the Morin transition (Methods).
For $T<T_{M}$, Fig. 1b, \FeO\ is expected to have mostly OOP spins
and indeed we observe large OOP domains (purple regions) separated
by AFM anti-phase domain walls $-$ ADWs (yellow boundaries). Although
the time-reversed OOP domains produce the same contrast in XMLD, the
magnetic sublattices are expected to be reversed by 180° on either
sides of the ADWs, so that spins rotate OOP$\rightarrow$IP$\rightarrow$OOP
across the yellow boundary (Extended Data Fig. 7). Upon warming to
$T\apprle T_{M}$, Fig. 1c-1g, the OOP anisotropy weakens causing
the ADWs (containing IP spins) to widen. This is accompanied by nucleation
of small IP islands (examples are encircled in green in Fig. 1c,1d)
emerging in the OOP matrix (purple). Above the transition, $T>T_{M}$,
IP domains dramatically enlarge to become the matrix (yellow regions,
Fig. 1h-1l), while the remaining OOP islands gradually shrink to become
fine bubbles (small purple dots), which persist up to room temperature.
The overall evolution of the AFM textures seen in Fig. 1b-1l can be
reversed by cooling the sample across $T_{M}$ (Extended Data Fig.
4), consistent with a hysteresis of $\sim$15 K (seen in the magnetometry
results, Fig. 1a). Likewise, at the \RhFeO$-$Pt interface, we find
very similar AFM textures consistent with $T_{M}$ being at room temperature,
Extended Data Fig. 3. We also collected images using circular dichroism
(XMCD-PEEM, see Methods) which are completely featureless (Extended
Data Fig. 5), indicating that XMCD-PEEM is insensitive to the weak
ferromagnetism above $T_{M}$, and that the features observed in Fig.
1b-1l are purely of AFM origin.

\subsection*{Merons, Antimerons and Bimerons}

Néel vector maps of the the IP orientations of $\boldsymbol{L}$ (see
Methods) are displayd in Fig. 2a-2c, and are labeled using a red-green-blue
(R-G-B) color-scale with 180° periodicity, while OOP orientations
are shown as white. At $T<T_{M}$ (Fig. 2a), most regions have OOP-oriented
$\boldsymbol{L}$, while the ADWs display an intricate IP winding
(R-G-B colors). A detailed analysis reveals the presence of mixed
Bloch-/Néel-type domain wall regions (indicated as black and yellow
straight lines across the ADWs), whereby the $\boldsymbol{L}$ rotates
in a plane oriented anywhere between \emph{transverse} or \emph{longitudinal}
to the wall width, respectively (Extended Data Fig. 7). Since a strong
\emph{interfacial} DMI (iDMI) would produce spin textures with fixed
chirality \cite{Ferrimagnet_fastmotion,Domain_wal_Logic_Nature,Relativistic_motion_AFM_walls_PRL},
the variation of the chirality along the ADW length in our data is
a clear indication that the contribution of iDMI is much weaker than
the already weak basal anisotropy. Lastly, our analysis indicates
that the small IP islands nucleating inside the OOP matrix have random
IP orientations consistent with the first-order nature of the Morin
transition.

When $T\sim T_{M}$, the OOP regions shrink (Fig. 2b), while the ADWs
and IP islands widen significantly, sometimes merging with each other
if in immediate proximity. When $T>T_{M}$ (Fig. 2c), we observe complex
AFM textures in which spins are predominantly lying in the basal planes,
separated by 60° domain walls \cite{Francis_Fe2O3_vortex}. When a
region of OOP spins (purple dots, Fig. 2d) happens to be encircled
by IP spins with a non-trivial topological Néel\emph{ }winding number
(\emph{i.e.}, $w\ne0$, demarcated as circles or squares in Fig. 2c),
the OOP bubble becomes \emph{trapped} and shrinks to a very small
size, but does not disappear, since it is topologically forbidden
from unwinding completely into the plane. At these topological pinch-points,
the domains merge to form two types of 6-fold pin-wheels (R-G{*}-B-R{*}-G-B{*}
and R-B{*}-G-R{*}-B-G{*}, asterisk corresponding to time-reversal)
\cite{Francis_Fe2O3_vortex}. Around the center of the pinwheels,
$\boldsymbol{L}$ circulates IP either along or against the azimuthal
angle, consistent with topologically distinct vortices (\emph{w} $=+1$)
and anti-vortices (\emph{w} $=-1$). Here, LV-PEEM image (Fig. 2d)
measured at the same position as the vector map (Fig. 2c) clearly
indicates the presence of OOP bubbles trapped at the (anti)vortex
cores. This implies that the textures observed in our samples at $T>T_{M}$
are actually the hitherto unreported AFM\emph{ half-skyrmions} (\emph{i.e.}
merons, antimerons), containing an OOP AFM core surrounded by whirling
IP AFM vortices and antivortices, see schematics in Fig. 2e. The Néel
topological charge of the AFM (anti)merons (\emph{Q} $=\pm1/2$) depends
on both the winding number and the core orientation, so that a core-up
meron is topologically equivalent (same \emph{Q}) to a core-down antimeron,
see Extended Data Fig. 8,9.

The magnetic evolution captured in Fig. 1,2 as a function of temperature
is highly reminiscent of the celebrated Kibble-Zurek transition \cite{Kibble1976,Zurek1985},
whereby a continuous symmetry is broken across a second-order phase
transition. We must, however, point out three key differences: firstly,
our system possesses discrete rather than continuous symmetries, though
the basal anisotropy breaking the $U(1)$ symmetry is quite weak.
Secondly, the Morin transition is first-order, and the formation of
AFM topological textures is governed by nucleation rather than fluctuations.
Thirdly, here the topologically trivial phase is \emph{ordered} with
a different spin direction, leading to a wide variety of 3-dimensional
spin textures. As we previously remarked, the process is completely
reversible: \emph{i.e.} the (anti)merons can be destroyed by cooling
(Extended Data Fig. 4) or by application of a magnetic field (discussed
below), and recreated by re-warming across $T_{M}$. After a complete
thermal cycle, most topological textures appear at completely new
positions. Although some persist at the same location from one cycle
to the next, deterministic reproduction over several cycles is rare,
suggesting that pinning plays some role but is not a dominant factor
(Extended Data Fig. 6).

By analyzing the pattern of AFM Néel vectors in Fig. 2c, we can recognise
several variants of the two basic topological types (Fig. 2e). We
observe both Néel-type (`hedgehog') and Bloch-type (`vortex') merons,
as well as antimerons, which combine these characteristics in their
different sectors. Of further interest is the presence of numerous
AFM meron-antimeron pairs with some bound very closely ($\sim$100-150
nm), and some others with their cores further apart (Fig. 2c, 2e).
In all these cases, the total topological charge of a pair is an integer:
a meron and an antimeron with the \emph{same} $Q$ give rise to a
$Q=\pm1$ AFM \emph{bimeron}, another hitherto unobserved quasi-particle
that is topologically equivalent to skyrmions \cite{Bimerons_1,Mostovoy_fractionalized_merons,Current_driven_motion_AFM_bimerons_PRL,asymmetric_bimeron_in_plane}.
In contrast, a meron and an antimeron with \emph{opposite} $Q$ form
a topologically trivial meron pair (TTMP) (with total $Q=0)$. TTMPs
do not enjoy strict topological protection, but could still form a
metastable bound state with a net $\boldsymbol{L}$ at its core. Crucially,
the total winding number away from the cores is \emph{zero} for both
bimerons and TTMPs, so that like skyrmions but \emph{unlike} isolated
(anti)merons, they could exist in a uniform AFM background (see Extended
Data Fig. 8,9) --- another important feature for constructing race-tracks.
Although we cannot determine the total topological charge of meron
pairs because of the sign ambiguity of $Q$, we observe that (anti)merons
which had formed separately, but were accidentally located nearby,
can either undergo annihilation (black ellipses, Fig. 1h,1i), or may
come closer together and remain stable (red ellipses in Fig. 1j-1l
and dashed-black ellipses in Fig. 2c). The former could be TTMPs,
which are not fully topologically protected, while the latter may
be \emph{Q} $=\pm1$ bimeron candidates. Interestingly, we never observed
the disappearance of \emph{individual} (anti)merons even in our longest
experimental timescales ($\sim$ 15 h at 300 K), indicating that the
lifetime of our AFM (anti)merons is several orders of magnitude longer
than predicted for AFM skyrmions \cite{Lifetime_stability_AFM_skyrmions}.

\subsection*{Control of texture dimensions}

We find that the feature sizes of the AFM textures (\emph{i.e.}, ADW
widths and (anti)meron cores) are strongly temperature dependent.
These sizes are controlled by the competition between the exchange
stiffness ($A_{ex}$) and the anisotropy ($K$) (known from the literature
\cite{Besser_dopedFe2O3,Coey_Rh-doped_Fe2O3}), and are expected to
diverge when $\left|K\right|\rightarrow0$. Based on the simple analytical
ansatz of linear AFM ADWs and linear (anti)merons, we obtained the
following expressions for the ADW width ($W$, when $T<T_{M}$), and
the (anti)meron cores ($D$, when $T>T_{M}$), see Supplementary Section
2 and 3,

\begin{equation}
\begin{array}{c}
W=\sqrt{\dfrac{2\pi^{2}A_{ex}}{K}},\\
\\
D=\dfrac{4}{3}\sqrt{\dfrac{2\pi^{2}}{(\pi^{2}-4)}\dfrac{A_{ex}}{\left|K\right|}}.
\end{array}
\end{equation}
In the absence of iDMI, these formulae hold irrespective of the textures'
Bloch- or Néel-character. The results of these models are in good
agreement with experimental data extracted from PEEM for both \FeO$-$Pt
and \RhFeO$-$Pt interfaces (see Fig. 3). Near $T_{M}$, both $W$
and $D$ diverge as $\left|K\right|\sim0$. However, away from $T_{M}$,
as the magnitude of $K$ increases, both $W$ and $D$ decrease asymptotically.
This demonstrates that it is possible to tune the room-temperature
size of (anti)meron cores by either reducing $A_{ex}$ or instead
increasing $\left|K\right|$, noting that the latter would also reduce
$T_{M}$. Furthermore, the minimal intra-bimeron core-core distance
should also be controllable by similar means. The possibility of manipulating
meron core sizes either by chemical doping or by exploiting the additional
anisotropy induced by an over-layer is of significant interest for
practical applications.

\subsection*{Annihilation of topological textures}

Lastly, we study the effect of IP magnetic fields, applied \emph{ex
situ}, and imaged the sample in remanence by LV-PEEM (Fig. 4). The
initial zero-field state at room temperature had an (anti)meron number
density of $n=$ 0.45 $\pm$ 0.07 $\mu$m\textsuperscript{-2}, which
remains unaffected within error ($\sim$ 0.41 $\mu$m\textsuperscript{-2})
after application of a 50 mT field, Fig. 4b. Errors are experimental
s.d. obtained from multiple images. However, after applying 500 mT,
we found large-scale annihilation of the (anti)merons, with very few
surviving textures ($n=$ 0.01 $\pm$ 0.02 $\mu$m\textsuperscript{-2}),
Fig. 4c. This field `erasability' may result from the movement of
the IP domains under the influence of the Zeeman coupling between
the field and $\boldsymbol{m}$ (see Extended Data Fig. 2). After
field erasure, the topological textures were re-generated by thermally
cycling the sample across $T_{M}$, Fig. 4d. In this state $n=$ 0.43
$\pm$ 0.04 $\mu$m\textsuperscript{-2}, which is close to the original
density. Crucially, the AFM topological textures are regenerated throughout
the sample, as confirmed by our large-area measurements (Extended
Data Fig. 6). Hence, in \FeO\ and its derivative systems, the Kibble-Zurek-like
phenomenology occurring across $T_{M}$ can be used as a reversible
handle for annihilation and re-generation of AFM (anti)merons and
bimerons.

\subsection*{Outlook}

Our results raise an important question: given that currents in the
Pt layer are expected to result in spin accumulation at the \FeO$-$Pt
interface via spin Hall effect, can some members of the AFM topological
family be manipulated via SOTs? Recent results \cite{SHE_Fe2O3_Pt_1_SOT_strain,SHE_Fe2O3_Pt_2_SOT}
of IP domain switching in \FeO\ suggest that this may become possible.
When realised in practice, such SOT-driven AFM (anti)meron/bimeron
motion would open up intriguing possibilities. For example, current-based
relativistic motion of these textures may serve as source of nanoscale
high-frequency electro-magnetic radiation \cite{Relativistic_motion_AFM_walls_PRL,Current_driven_motion_AFM_bimerons_PRL,Galkina}.
Alternatively, these textures could be generated in Rh-doped \FeO\ at
room temperature by Kibble-Zurek cycling or possibly by optical \cite{light_spin_reorientation_TmFeO3,ultrafast_generation_skyrmions_AFMI}
or electrical \cite{Current_driven_motion_AFM_bimerons_PRL} stimuli,
and be driven electrically via SOTs, serving as information vectors
in race-track-based memory and logic implementations \cite{2020_roadmap,AFM_skyrmion_Logic_gates,Neuromorphic_spintronics,Skyrmions_application,Skyrmion_electronics}.

\pagebreak{}

\bibliographystyle{naturemag}
\bibliography{AFMmeron}

\section*{Supplementary Materials}

\noindent 1. Materials, Methods and Notes: Growth, magnetic and X-ray
characterization, XMLD- and XMCD-PEEM imaging, generation of Néel
vector maps, definition of topological numbers.

\noindent 2. Extended Data: (1) Characterization of films, (2) Field-dependent
magnetometry, (3) AFM textures in Rh-doped \FeO, (4) AFM textures
in \FeO\ during cooling, (5) XMCD-PEEM, (6) Large scale reproduction
of topological textures and role of pinning, (6) AFM Néel and Bloch
ADWs, (7) AFM Néel (anti)merons and bimerons, (8) AFM Bloch (anti)merons
and bimerons. 

\noindent 3. Supplementary Information: Angular dependence of X-ray
absorption. Magnetic modelling of ADW width and (anti)meron core size.
Exchange interaction terms in \FeO\ to calculate exchange stiffness.

\section*{Acknowledgments}

We would like to thank S. Parameswaran for discussions, F. P. Chmiel
for guidance with data reduction, and R. D. Johnson for assistance
with experiments. We acknowledge Diamond Light Source for time on
Beam Line I06 under Proposals MM23857 and S120317. The work done at
University of Oxford (H.J., J.-C.L., J-H.C., J.H. and P.G.R.) is funded
by EPSRC grant no. EP/M2020517/1 (Quantum Materials Platform Grant).
The work at National University of Singapore (H.J., S.P., A.A. and
T.V.) is supported by National Research Foundation (NRF) under Competitive
Research Program (NRF2015NRF-CRP001-015). A.A. would like to thank
the Agency for Science, Technology and Research (A{*}STAR) under its
Advanced Manufacturing and Engineering (AME) Individual Research Grant
(IRG) (A1983c0034) for financial support. The work at University of
Wisconsin-Madison (J.S. and C.B.E.) is supported by the Army Research
Office through Grant W911NF-17-1-0462 and the National Science Foundation
under DMREF Grant DMR-1629270.

\section*{Author Contributions}

H.J. performed material optimization, thin film growth by PLD, structural
and magnetic characterization. H.J., J.-C.L., J.H., F.M. performed
PEEM experiments. J-H.C., J.-C.L., H.J. performed data reduction and
analysis. J.-C.L., S.P. performed over-layer growth and surface characterization.
J.-C.L. assisted in magnetic characterization. J.S. prepared preliminary
sputter-grown samples under the supervision of C.-B.E. A.A. and T.V.
supervised the PLD film growth and characterization. H.J., under the
guidance of P.G.R., prepared the theoretical model. P.G.R. and H.J.
conceived the project and supervised the analysis. H.J. and P.G.R.
prepared the first draft of the manuscript. All authors discussed
and contributed to the manuscript. Requests for supplementary materials
and correspondence should be addressed to H.J. or P.G.R. 

\pagebreak{}

\section*{Main Figure Legends}
\begin{center}
\includegraphics[scale=0.95]{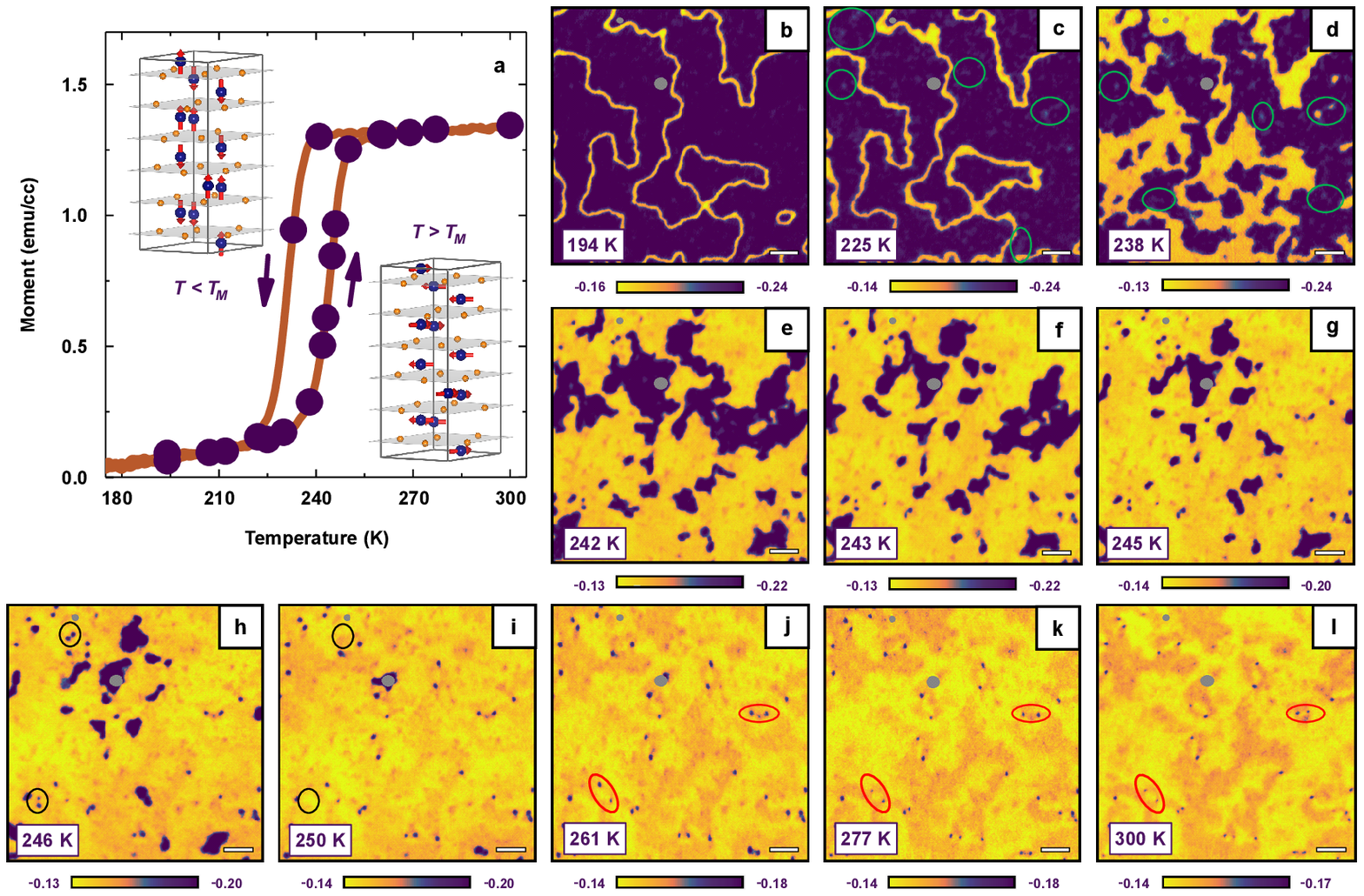}
\par\end{center}

\textbf{Figure 1. Temperature evolution of AFM textures across }$\boldsymbol{T}_{\boldsymbol{M}}$\textbf{
in} \textbf{$\boldsymbol{\alpha}$}-\textbf{Fe}$_{\boldsymbol{2}}$\textbf{O}$_{\boldsymbol{3}}$\textbf{.}
(\textbf{a}) Magnetometry data (brown) of \FeO\ showing hysteresis
of the IP moment ($\boldsymbol{m}$) across the Morin transition,
with purple circles indicating PEEM measurement temperatures. Insets
show the Fe-spin orientations in the hexagonal unit cell below/above
$T_{M}$. (\textbf{b-l}) AFM textures in LV-PEEM vs temperature at
the \FeO$-$Pt interface, on warming across $T_{M}$ (cooling sequence
in Extended Data Fig. 4. Overall evolution in Supplementary Video
S1). Yellow and purple contrast regions indicate IP and OOP AFM orientations,
respectively. Fine purple dots (in h-l) are OOP bubbles, later shown
(Fig. 2) to be lying at the core of AFM (anti)merons. All images were
recorded at the same position. Energy contrast scales (a.u.), indicated
adjacent to each image, were varied slightly across the transition
to provide better contrast (see Methods). Spatial scale bars are 1
$\mu$m. Green ellipses (in c,d) indicate some examples of IP islands
in the OOP matrix emerging near $T_{M}$. Black ellipses (in h,i)
highlight the meron/antimeron pairs that undergo annihilation upon
warming. Red ellipses (in j-l) encircle stable meron/antimeron pairs
that come closer upon warming. Grey regions (in b-l) are defects on
the sample surface to focus the PEEM image (Methods).
\begin{center}
\includegraphics[scale=0.95]{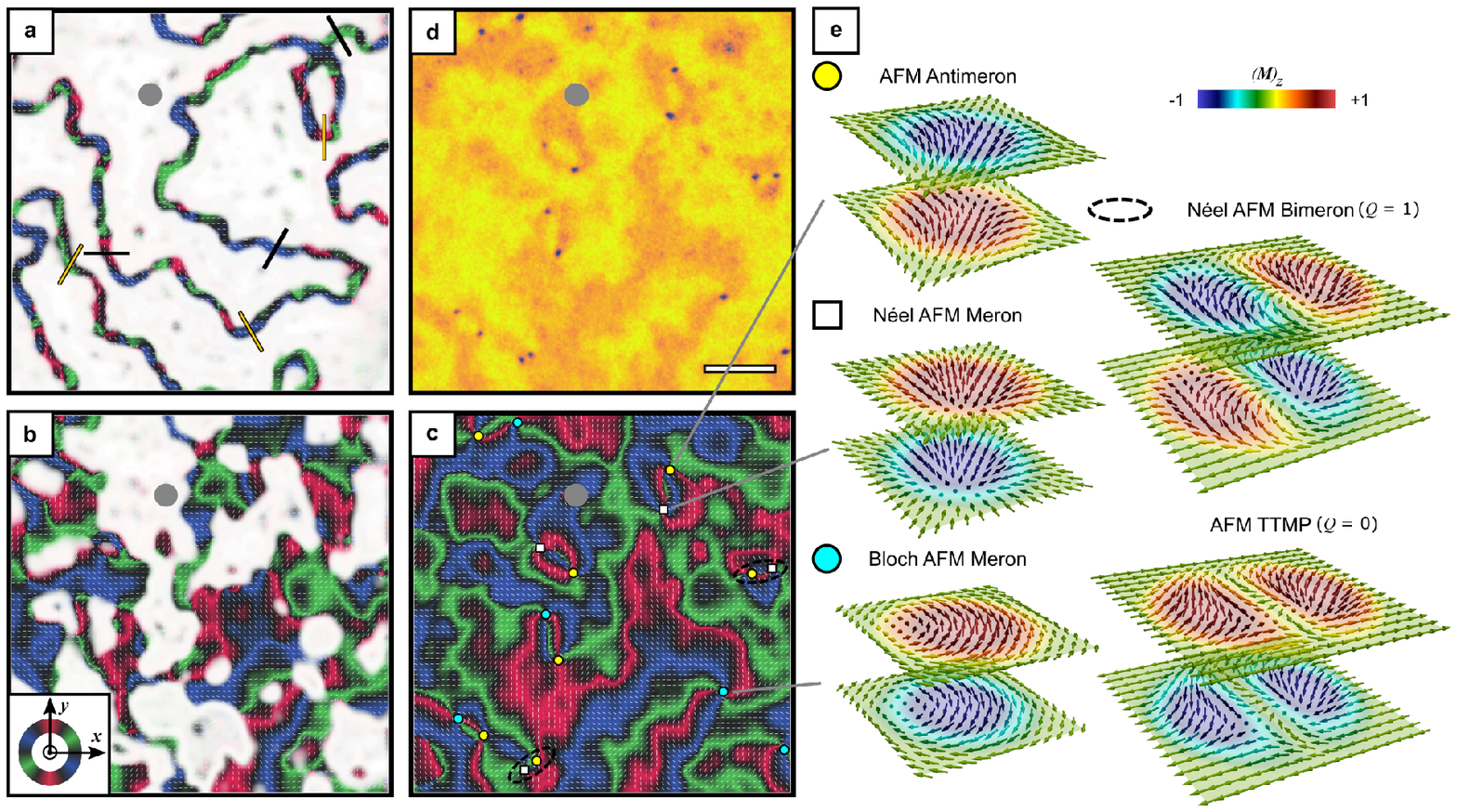}
\par\end{center}

\textbf{Figure 2. Real-space topological textures from in plane Néel
vector maps.} Vector-mapped LH-PEEM images at the \FeO$-$Pt interface,
as described in Methods, at (\textbf{a}) $T<T_{M}$, (\textbf{b})
$T\sim T_{M}$, and (\textbf{c}) $T>T_{M}$. R-G-B colors (legend
in (b) inset) and thin white bars represent IP $\boldsymbol{L}$ orientations.
White regions represent OOP $\boldsymbol{L}$ orientations, and black
regions highlight IP $\boldsymbol{L}$ directions deviating significantly
from the R-G-B directions. Black and yellow straight lines in (a)
demarcate ADW cross-sections with predominantly Bloch- and Néel-type
characters, respectively. (\textbf{d}) LV-PEEM image at the same conditions
as in (c) and Fig. 1l, showing AFM OOP bubbles lying at the core of
the corresponding (anti)vortices in (c). Spatial scale bar is 1 $\mu$m.
All images were recorded at the same sample position. (\textbf{e})
Schematics of the AFM topological textures demarcated in (c) as circles
or squares. The AFM topological family (including variants related
by core reversal) is shown in Extended Data Fig. 8,9. The color legend
in (e) corresponds to the OOP spin component of the two magnetic sublattices.

\pagebreak{}
\begin{center}
\includegraphics[scale=0.95]{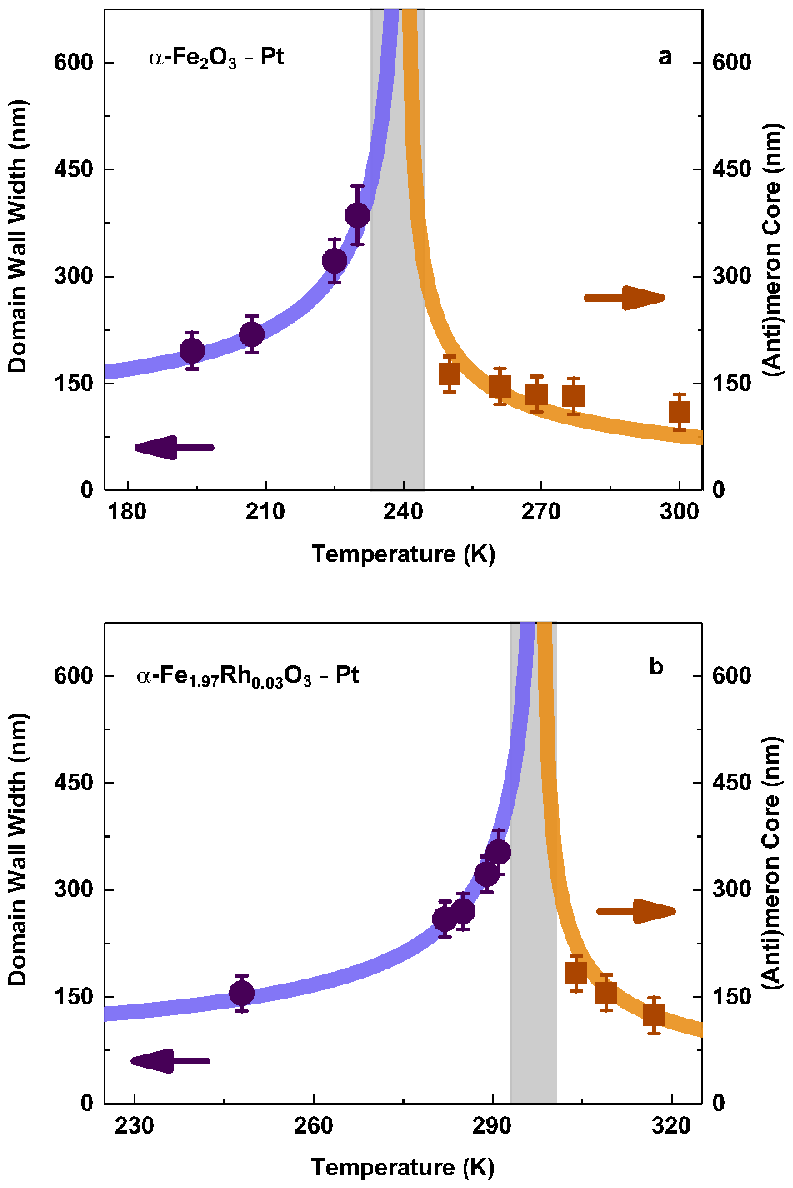}
\par\end{center}

\textbf{Figure 3. Temperature evolution of AFM feature sizes in (a)
$\boldsymbol{\alpha}$}-\textbf{Fe}$_{\boldsymbol{2}}$\textbf{O}$_{\boldsymbol{3}}$\textbf{
and (b) $\boldsymbol{\alpha}$}-\textbf{Fe}$_{\boldsymbol{1.97}}$\textbf{Rh$\boldsymbol{_{0.03}}$O}$_{\boldsymbol{3}}$\textbf{.}
The purple and orange lines are the calculated AFM texture dimensions,
$W$ and $D$, as a function of temperature based on our analytical
models (see Supplementary sections 2 and 3). Purple circles and orange
squares represent corresponding averaged experimental results of ADW
widths and (anti)meron core sizes, respectively, extracted from the
LV-PEEM images (see Methods). Error bars represent either experimental
s.d. or PEEM resolution limit, whichever is larger. The temperature
zones demarcated as grey are in the proximity of the Morin transition
where we do not provide experimental data points since the definition
of ADWs and (anti)merons breaks down and all structures become very
large, see Eqn (1).

\pagebreak{}
\begin{center}
\includegraphics[scale=0.95]{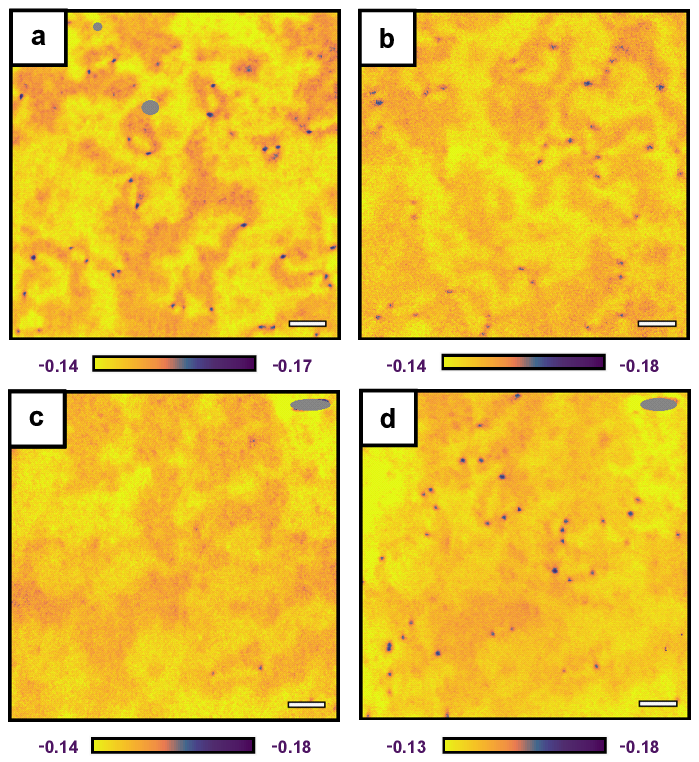}
\par\end{center}

\textbf{Figure 4. Erasing and re-creating (anti)merons.} Room temperature
LV-PEEM images at \FeO$-$Pt interface in (\textbf{a}) its original
state, and in its remanent state after application of (\textbf{b})
50 mT and (c) 500 mT \emph{ex situ} IP magnetic fields. (\textbf{d})
Final state after the sample in state (c) was thermally cycled (cooling
then warming) across $T_{M}$. Images (a,b,c) were recorded in different
locations on the sample, whereas images (c,d) were recorded at the
same position. Spatial scale bars are 1 $\mu$m. 
\end{document}